# Applying AOSE Concepts to Model Crosscutting Variability in Variant-Rich Processes

Tomás Martínez-Ruiz, Félix García, Mario Piattini
Alarcos Research Group, Department of Information Technologies and Systems, Escuela Superior de Informática, University of Castilla - La Mancha
Paseo de la Universidad, 4 13071 Ciudad Real, Spain
+34 926 295300 ext. 6648, 6475, 3715
{tomas.martinez, felix.garcia, mario.piattini}@uclm.es

Jürgen Münch
University of Helsinki and Fraunhofer IESE
Gustaf Hällströmin katu 2b, FI-00014 Helsinki, Finland
+358 9 191 51330
juergen.muench[at]cs[dot]Helsinki[dot]fi

*Abstract*— Software process models need to be variant-rich, in the sense that they should be systematically customizable to specific project goals and project environments. It is currently very difficult to model Variant-Rich Process (VRP) because variability mechanisms are largely missing in modern process modeling languages. Variability mechanisms from other domains, such as programming languages, might be suitable for the representation of variability and could be adapted to the modeling of software processes. Mechanisms from Software Product Line Engineering (SPLE) and concepts from Aspect-Oriented Software Engineering (AOSE) show particular promise when modeling variability. This paper presents an approach that integrates variability concepts from SPLE and AOSE in the design of a VRP approach for the systematic support of tailoring in software processes. This approach has also been implemented in SPEM, resulting in the vSPEM notation. It has been used in a pilot application, which indicates that our approach based on AOSE can make process tailoring easier and more productive.

*AOSE, Variant-Rich Processes, Process Variability, Process Tailoring, Process Lines, Aspect Oriented Software Development.*

## I. INTRODUCTION

The tailoring of software processes makes them fit for different organizations and projects [1]. It also leads to the appearance of certain needs that are similar to those found in Software Product Line Engineering (SPLE). The Variant-Rich Process (VRP) approach was developed to meet these needs. This approach proposes the inclusion of variants of process-composing elements in the processes themselves through the use of on-point variations, thus making process tailoring easier. It is based on applying assets from SPLE to processes [2], since software products and processes have commonalities [3].

However, these variations are too detailed for industries, which need to tailor processes through crosscutting variations [4], including several synchronized on-point variations. These crosscutting variations may consist of including certain characteristics or criteria, e.g. security, within a process model, without taking into account its implementation: *which* variants need to be used, or *how*. Crosscutting variations in software processes are still not well managed by the sole application of SPLE assets. This being so, Variant-Rich Process variations could be considered as concerns, and could thus be managed with Aspect-Oriented Software Engineering (AOSE)-based variability mechanisms when they are crosscutting. The "weaving" of these variations leads to the tailored process.

Aspect-Oriented Software Engineering [5] provides the capability with which to identify and encapsulate the crosscutting concerns within aspects, which can then be more consistently managed. Literature shows that aspects have been used to build software processes [6, 7] and to manage process variability [8]. In this work we therefore propose extending the existing SPLE-based Variant-Rich Process approach with AOSE-based crosscutting variations and aspectual weaving. The vSPEM [9, 10] notation has also been enriched with new AOSE-inspired constructors. This is thus the basis for supporting process tailoring as required in real processes [4].

This paper is organized as follows. Section II presents the state of the art. Section III describes the extension of the vSPEM proposal with AOSE-based variability. A pilot study, and lessons learned are shown in Section IV. Section V deals with our conclusions and future work.

## II. STATE OF THE ART

The behavior of a software process depends on its structure [11], and process tailoring activities are useful in making each process fit its appropriate behavior when it is instantiated. The topic of process tailoring is widely dealt with in literature, as Pedreira et al. show [12], and is supported in several ways, as the systematic review of Martinez-Ruiz et al. demonstrates [4]. Tailoring is carried out by means of changes to the elements of which the processes are composed and to their relations, principally in activities, artifacts, and roles. Variations are also executed in two ways, namely in a detailed and intensive manner in order to configure each element with its suitable value, and in a crosscutting manner in order to configure a large number of process composing elements simultaneously.

Moreover, since the emergence of the product line approach [13], most of the initiatives with which to model variability in processes have attempted to adapt it [4], as was proposed by Rombach in [2]. Some works concerning process tailoring have thus been proposed based on process lines and variation mechanisms [14], using stereotypes [15] and defining new specific notations [9, 10]. Other proposals include environments in which to represent variability and to carry out the tailoring and monitoring of the process during its execution

[16, 17]. According to [18], tailoring based on Variant-Rich Processes is used to support process institutionalization .

Some approaches deal with the application of AOSE to product lines. According to Kulesza et al. [19], aspects may be used to introduce flexibility into a system. Each aspect signifies temporal behavior, which may imply variability. Laddaga et al. [20] show aspect management techniques that may be applied to dynamic architectures, mainly product lines. In this respect, Apel et al. [21] present some problems of Feature-Oriented Programming and how they are solved by means of using AOSE. The correspondence between elements from AOSE and variability concepts in product lines is shown in [22] and [23]. González-Bauxili et al. [24] present a meta-model that can be used to combine aspects from product lines and aspects, and they define how to transform traditional models into aspect models.

According to López Herrejon and Batory [25], advice and pointcuts may modify a class or method behavior. They can also be formalized by using use case slides and algebra techniques. Colyer et al. [26] describe some of the limitations involved in applying aspects to product lines: the concerns must be orthogonal, without dependencies between them, in order to allow their independent use. Mezini and Ostermann [27] present an analysis of how to model variability using mechanisms of aspect-oriented languages, mainly Caesar. Figueiredo et al. [28] present and evaluate the suitability of AOSE in modeling product lines and compare the results with other product line approaches. The results obtained show that AOSE is more efficient.

Aspects have also been combined with business processes, as proposed by Odgers [29], because of their flexibility and dynamism. Moreover, Charfi et al. [30] present AO4BPEL, an aspect-oriented extension to BPEL which deals with support for crosscutting concerns and dynamic adaptation. In addition, the work of Sutton [8] allows us to consider Aspects in processes [31]. Quites el al. [7] propose using them to design high-level management policies in software process models. Mishali and Katz [6] propose applying aspects to monitor or enforce XP practices over Eclipse.

In a previous work, vSPEM, a SPLE-based Variant-Rich Process language which supports the realization of punctual variations over the processes, was proposed [9]. This proposal is suitable when specific variations have to been applied to process models, but real adaptations found in literature show that processes are sometimes tailored by using *crosscutting* variations, which affect more than one element each time [4]. Some on-point variations must therefore be included and the consistence between them guaranteed. The use of only punctual variations may make this task tedious, as the process engineer must enter the process structure and seek the variation points that satisfy all the on-point variations according to a particular criterion. This therefore forces him/her to focus on how the variations are carried out and where they are executed, rather than abstracting from it and focusing on the business requirements that motivate the variation.

Since crosscutting variations are not suitably supported in vSPEM, it has been enriched with variability mechanisms based on AOSE. The main advantage of the proposal presented in this paper with regard to others found in literature is that it supports process tailoring based on aspect mechanisms rather than being focused solely on process reuse. It is also a generic approach, which could be used to carry out any variation (including any specific characteristics) in a software process, and in any of the process constructors.

## III. ENRICHING THE VARIANT-RICH PROCESS APPROACH WITH AOSE CONCEPTS

The AOSE approach was analyzed and mapped onto the SPLE-based Variant-Rich Process approach through an analysis of the AspectJ language [32, 33] and its constructors (see references [32, 33] for further information about these constructors). It was then enhanced with the capability to handle crosscutting variations. The vSPEM notation was also enriched with new AOSE-inspired constructors. The added or modified elements are highlighted in Fig. 1.

### A. Crosscutting Variability in Variant-Rich Processes

In order to introduce aspects into process line variability, analogies may be determined between the concepts in AOSE and Variant-Rich software processes, as Tables I and II show.

Both tables show that there are similar concepts in software programming and software processes. They also show that variant rich processes lack some of those which could be used to manage crosscutting variations by using the AOSE paradigm assets. These have been then modeled as Fig. 1 presents.

### B. Variation Points (Redefined)

Variation points are the places where the elements included in the crosscutting variation should be inserted. Table III describes the different types of variation points that exist in Variant-Rich Processes, based on analyzing the process elements' behavior and their variability support.

TABLE I. MAPPING BETWEEN AOSE AND PROCESS CONCEPTS

| Programming Concept | Variant-Rich Process Concept |
|---|---|
| Class | Process or sub process |
| Methods | Work Units (activities, tasks …) |
| Attributes | Work Products |
| Constants | Resources (humans, tools…) |

TABLE II. COMPARISON BETWEEN THE AOSE AND VRP CONCEPTS

| AOSE Concept | New Variant-Rich Process Concept |
|---|---|
| Crosscutting concern | Crosscutting Variation |
| Aspect | Process Aspect (see Sect. III.E) |
| Join point | Variation Point (see Sect. III.B) |
| Pointcut | Process Pointcut (see Sect. III.C) |
| Advice declaration | Process Advice (see Sect. III.D) |
| Behavior in the advice | Variants |

TABLE III. VARIATION POINTS EXISTING IN A PROCESS DEFINITION

| AOSE Join Point | Variant-Rich Process Variation Point |
|---|---|
| A call to a method or constructor | Call of a work unit (activity, task) the called element (in the caller element) |
| Execution of method, constructor or advice | Call of a work unit (activity, task…) in the called element. |
| Access or update of a field | Use or creation of a work product |
| Access of a field | Use of a resource (human, tool) |
| Initialization of classes and objects | First use or delivery of work products |

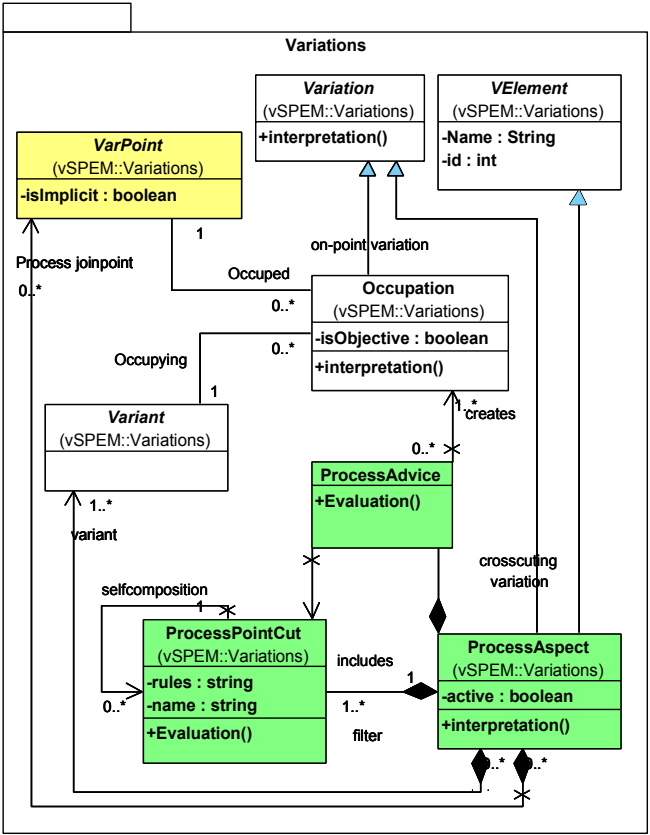

Figure 1. Elements for managing crosscutting variations

As is described above, variation points are explicitly created places into which variants contained in on-point variations can be inserted. It is necessary to differentiate between the variation points created for realizing on-point variations (based on SPLE), and all the existing and well-known variation points used in executing crosscutting variations. A new attribute (*isImplicit*) was therefore added to the *VarPoint* element which indicates whether the variation point has been created for on-point and crosscutting variations (*explicit*) or is obtained only for crosscutting ones (*implicit*).

*C. Process Pointcut*

Process Pointcuts determine which variation points are used to introduce the on-point variations. Determination takes place automatically once the process aspect has been *activated*. All the variation points in the process are analyzed in order to select those that are to be occupied, and, if necessary, to choose some variants with which to extract the context. Similarly to pointcuts (in AOSE), process pointcuts are defined as:

*Pointcut* <*name*> (<*VariabilityElements list*>) : <*expressions*>

The *expressions* are built by means of the designators[1] and logic connectors (&&, || and !). They may refer to other process pointcuts. Pointcuts filter certain variation points (and sometimes variants), which are given to the process aspect to make the corresponding on-point variations.

---

[1] Complete list of designators is in http://alarcos.esi.uclm.es /per/tmartinez/vSPEM/aspectConstructors.html

The properties of Process Pointcuts are their names and expressions. The latter is a string which filters the variation points used to carry out the crosscutting variation. Pointcuts also include a method with which to evaluate the expression and obtain the variability elements that satisfy this expression.

*D. Process Advices and Variants*

Process advices, therefore, focus on modifying the occupation relationships between the variation points and the variants to satisfy the crosscutting variation. This is achieved by creating certain designators (see Note 1). By using these, *process advices* are declared as:

*Advice* <*process pointcut*> *(*<*VariabilityElements list*>*) {* <*actions of the crosscutting variation*> *}*

"*Advice*" is therefore the word used to indicate process advice; *pointcut* is the name of the process pointcut that activates the advice (or the expression combining process pointcuts by using logic connectors). The *evaluation* method creates occupation relationships by using the *variability elements*.

*E. Process Aspect*

Process aspects support every crosscutting variation in variant-rich software processes. They are therefore composed of process pointcuts and some process advices, and include some variants and perhaps variation points that are used to carry out the crosscutting variation. They also encapsulate and abstract the user from the aforementioned elements.

Since process aspects are used to introduce variability by carrying out variations, they are a specialization of both the *VElement* constructor (more details of *VElement* can be found in [9]). Process aspects also include the *active* attribute which shows whether the aspect is used to tailor the Variant-Rich Process, and one operation which makes the actual changes.

IV. PILOT STUDY USING CROSSCUTTING VARIATIONS

The aforementioned approach has been used to model variability in the Jaxa Process Line (more details about this process line can be found in [34]). This process line includes several variabilities (as Table IV details), but when it is tailoring, some of these are similarly treated, depending on the same criterion. The on-point variations of these points are encapsulated in one crosscutting variation. This study shows how to create a crosscutting variation for the first criterion.

This criterion implies that if Satellite 2 is developed, one task and one work product must be added in order to tailor the process. This implies one aspect, including some variants and variation points with which to realize the on-point variations.

The aspect additionally includes a process pointcut, which is also transparent from the end user's point of view. Its aim is to filter task and work product variation points in Activity *1.2.2., Software Design,* using the following constructors. Patterns have been used to avoid listing the full names:

```
pointcut ppc1 (VPTask vpt1, VPWorkP vpw1,
VPWorkP vpw2):
   vpt1=(execution("1.2.2*"));
   vpw1=(use(*) && within("1.2.2*"));
```

TABLE IV. LIST OF VARIABILITY ELEMENTS IN JAXA

| Variability | Criterion |
|---|---|
| FMECA work product | Include in satellite 2 |
| Analyze HW and SW interaction task | Include in satellite 2 |
| Rationale for each requirement artifact | Not included in science missions |
| Quality source code work product | Not included in science missions |
| Requirements in design work product | Not included in science missions |

This pointcut obtains a task variation point within the activity and may be used by the advice. Some other variation points are also retrieved. They contains the input and output products of the task. These variation points are taken by the advice. This is written as:

```
advice ppc1 (VPTask vpt1, VPWorkP vpw1,
VPWorkP vpw2){
   vpt1.occupe(Analyze HW SW Interact.);
   vpw1.occupe(FMECA);   }
```

The instructions in the process advice allow the variants (FMECA work product and the Analyze HW SW Interaction task) to be placed in the variation points obtained previously–*vpw1* and *vpt1*, respectively. Moreover, as is set out in detail in [10], the *variant2variant* dependency between them is converted into a relationship in the completely opposite direction. After interpreting variability, the *FMECA* work product and the *Analyze HW SW Interaction* task are then created as new elements in Activity *1.2.2., Software Design*.

With the AOSE-based mechanisms, the tailoring of a process with crosscutting variations is now simplified by activating aspects rather than selecting which variants to place in which variation points, and combining them. This reduces the effort and abstracts the user when tailoring the process.

*A. Lessons Learned*

After analyzing the application of the vSPEM in a real Variant-Rich Process, some advantages and disadvantages of using crosscutting variations were found. The first advantage was that crosscutting variations include several on-point variations that make it unnecessary to realize these (on point) variations manually, and this thus reduces the tailoring effort. Taking the pilot application we have described above, the Variant-Rich Process initially contained five on-point variations to be evaluated in order to obtain a tailored process, which signifies a tailoring effort of *5n*, where n is the number of tailored processes from the Variant-Rich process. After applying the approach proposed in this paper, these were reduced to two crosscutting variations. This signifies an effort of 2n, bringing about a reduction for half the initial effort for process tailoring. Once the AOSE-based mechanisms have been defined, they could be easily used and reused for tailoring processes. In addition, all the new elements defined over the process line are transparent from the user viewpoint, with the exception of the aspect, which is actually "activated". This signifies that the use of crosscutting mechanisms does not imply an increase in process tailoring complexity.

Moreover, when varying process composing elements, it is necessary to ensure that the variations of these elements remain consistent. The crosscutting of variations ensures that consistency is not the process engineer's responsibility, since it is configured in the process advices and process pointcuts, and allows the user to be abstracted out from such details, which is the second advantage. In addition, as process aspects are consistent themselves, they could be built without dependences between them, as Colyer [26] proposes.

However, AOSE-based crosscutting variations have one main disadvantage: they do not offer the user the ability to decide, in detail, which variants should occupy each variation point. On the other hand, this disadvantage of crosscutting variations is the most important advantage of SPLE-based on-point variations. As a result, the combination of both types of variations leads to the creation of the most complete process tailoring paradigm.

To sum up, this pilot study shows very promising results: Developing an SPLE&AOSE-based Variant-Rich Process approach is a promising and robust initiative for supporting effective process tailoring and making it feasible for organizations to use it to tailor their own software processes in the near future. In addition, the more complex the processes are, the more on-point variations they include, and the more are encapsulated into aspects, which makes them more useful mechanisms in tailoring real and complex software processes.

## V. CONCLUSIONS AND FUTURE WORK

Software processes must be tailored by means of on-point and crosscutting variations. The Variant-Rich Process approach must therefore focus on supporting both types of tailoring in order to offer organizations usable mechanisms with which to tailor their processes. Since there is no way to manage crosscutting variations through tailoring mechanisms based on product lines, this paper focuses on the transferred AOSE paradigm to fill this gap and to support process model tailoring as required by real processes.

As a result, the Variant-Rich Process approach has been enriched with new AOSE-inspired concepts such as the process aspect. These were merged with the previous SPLE-based concepts, with the intention of all of them being able to manage variability in a consistent and complementary manner.

Managing crosscutting variations in Variant-Rich Processes implies two main advantages. First, it ensures variability consistency, and second, it facilitates tailoring. As variations in real processes involve several elements, this reveals that crosscutting variability mechanisms are actually needed in order to guarantee consistency when varying all these elements. Moreover, tailoring processes from a Variant-Rich Process without consideration for crosscutting variations signifies that all on-point variations must be decided on one by one. In contrast, crosscutting variations allow abstraction, signifying that these variations can be carried out simultaneously.

Our approach has also been implemented on SPEM, resulting in the vSPEM notation. This notation has been used to model variability in a real process and later to resolve this variability by tailoring processes. The main lesson learned from the results obtained in the pilot study are that the AOSE-based Variant-Rich Process approach promises to be a powerful and robust initiative for modeling variability in software processes in alignment with tailoring requirements in current software development organizations. It also shows that it is possible not only to use AOSE concepts to model variability, but also to employ AOSE lifecycle techniques for scoping, along with determining variability in software process models –without variability.

As future work, we shall focus on analyzing other AOSE implementations to improve our approach. We also intend to carry out experiments to verify the usability of the vSPEM notation extended with AOSE, along with its applicability to real processes. This approach will also be included as a plug-in in the Eclipse framework in an effort to facilitate process tailoring. Finally, organizations need to tailor their processes before representing them, and if they are to improve, processes must be more and more capable. Process institutionalization is the only way to do this. The process tailoring techniques we have developed may therefore be considered as a basis for building an institutionalization framework based on process tailoring and standardization. As institutionalization leads towards better process evolution, fragility problems appearing in the Variant-Rich Process approach owing to the use of aspects will be mitigated. This will be achieved by controlling how the processes and their process aspects need to evolve.

ACKNOWLEDGMENTS

This work is partially supported by the Program FPU of the Spanish Ministerio de Educación, and by the PEGASO/MAGO (MICINN and FEDER, TIN2009-13718-C02-01), MEDUSAS (CDTI (MICINN), IDI-20090557), ALTAMIRA (JCCM, Fondo Social Europeo, PII2I09-0106-2463) and INGENIOSO (JCCM, PEII11-0025-9533), projects. We would also like to thank our colleagues at the Japanese Aerospace Exploration Agency (JAXA) for their collaboration, and Sonnhild Namingha from Fraunhofer IESE for reviewing a first version of this paper.